\documentclass[osajnl,twocolumn,showpacs,superscriptaddress,10pt]{revtex4-1}

\usepackage{amsmath,amssymb,graphicx}

\begin{document}

\title{Versatile photoacoustic spectrometer based on a mid-infrared pulsed optical parametric oscillator}

\author{Laurent Lamard}
\affiliation{Laserspec BVBA, 15 rue Trieux Scieurs, B-5020 Malonne, Belgium}

\author{David Balslev-Hardee}
\affiliation{Danish National Metrology Institute, Kogle Alle 5, 2970 H{\o}rsholm, Denmark}

\author{Andre peremans}
\affiliation{Laserspec BVBA, 15 rue Trieux Scieurs, B-5020 Malonne, Belgium}
\affiliation{University of Namur, Rue de Bruxelles 61, B-5000 Namur, Belgium} 

\author{Jan C. Petersen}
\affiliation{Danish National Metrology Institute, Kogle Alle 5, 2970 H{\o}rsholm, Denmark}

\author{Mikael Lassen}\email{Corresponding author: ml@dfm.dk}
\affiliation{Danish National Metrology Institute, Kogle Alle 5, 2970 H{\o}rsholm, Denmark}


\begin{abstract}
We demonstrate the usefulness of a nanosecond pulsed single-mode mid-infrared (MIR) optical parametric oscillator (OPO) for Photoacoustic (PA) spectroscopic measurements. The maximum wavelength ranges for the signal and idler are 1.4 $\mu$m to 1.7 $\mu$m and 2.8 $\mu$m to 4.6 $\mu$m, respectively, and with a MIR output power of up to 500 mW. Making the OPO useful for different spectroscopic PA trace-gas measurements targeting the major market opportunity of environmental monitoring and breath gas analysis. We perform spectroscopic measurements of methane (CH$_4$) nitrodioxide (NO$_2$) and ammonia (NH$_3$) in the 2.8 $\mu$m to 3.7 $\mu$m wavelength region. The measurements were conducted with a constant flow rate of 300 ml/min, thus demonstrating the suitability of the gas sensor for real time trace gas measurements. The acquired spectra are compared with data from the Hitran database and good agreement is found. Demonstrating a resolution bandwidth of 1.5 cm$^{−1}$. An Allan deviation analysis shows that the detection limit for methane at optimum integration time for the PA sensor is 8 ppbV (nmol/mol) at 105 seconds of integration time, corresponding to a normalized noise equivalent absorption (NNEA) coefficient of 2.9$\times 10^{-7}$ W cm$^{-1}$ Hz$^{-1/2}$. 
\end{abstract}

\maketitle

\section{Introduction}

For highly sensitive and selective photoacoustic (PA) trace-gas sensing it is desirable to have high power light sources with large wavelength tunability in the mid-infrared (MIR) region, where most molecules have strong vibrational transitions \cite{Hodgkinson2013,Sigrist2008,Nagele2000,Tomberg2018}. There are a number of different MIR light sources and technologies available, including gas lasers, distributed feedback diode lasers, coherent sources based on nonlinear optics, difference frequency generation and optical parametric oscillators (OPOs), and relatively new devices such as quantum and interband cascade lasers (QCL/ICL) are attractive devices. QCLs represent an excellent choice for PA sensing with high resolution and sensitivity \cite{Pushkarsky2006,Lassen2017}. However, QCL/ICL sources are still relative expensive and do not all access the atmospheric windows. Even though the OPO technology is an old technology it is still an excellent choice as light source for PA sensors \cite{Hirschmann2013,Peltola2013,Costopoulos2002,Arslanov2012,lassen2016OL,Hellstrom2013}. They provide molecular selectivity due to large wavelength tunability, high energy, and cost-effective and compact device for the generation of infrared light in the 1.5 to 5 $\mu$m spectral range \cite{Petrov2015}. OPOs also allow for better multigas detection of several components  with common signal processing and data analysis. MIR OPOs can operate in different configurations both pulsed and continuous wave (CW). The pulsed operation results in high peak power, nonlinear absorption effect might be an issue and the normalized noise equivalent absorption never reached extreme levels. However, pulsed OPO systems have wide tuning ranges and can is thus suitable for measuring most of the interesting trace gases. CW OPOs provide smaller wavelength tuning range, however, they offer higher sensitivity and in most cases higher spectral resolution \cite{Bartlome2009}.

We demonstrate a novel miniaturized PA trace gas analyser platform integrated with a MIR OPO targeting the major market opportunity of environmental monitoring \cite{Sigrist2004} and breath analysis \cite{Loureno2014,Saalberg2017}. The novel design of the MIR OPO differs from the low power doubling resonant scheme proposed in the work by \cite{Berrou2018} because it enables continuously tuning from 2.8 $\mu$m to 4.6 $\mu$m in 15-20 minutes with steps of 0.5 pm with high output power and in a single longitudinal mode (SLM). PAS sensors with pulsed MIR OPOs have been report in the literature before, however these OPOs have limited tuning range and lower output powers compared to the MIR OPO described here. It is specifically shown that a miniaturized photoacoustic spectroscopic (PAS) cell can be excited resonantly with the MIR OPO by adjusting the laser pulse repetition rate to match the frequency of the acoustic resonance of the cell. The application of the gas sensor for real time environmental measurements and breath analysis is demonstrated using three samples of gas concentration; 100 ppmV of methane (CH$_4$), 100 ppmV of nitrogen dioxide (NO$_2$), and approximately 1000 ppmV ammonia (NH$_3$) in atmospherical air with a humidity of 40$\%$. A gas flow rate of 300 ml/min through the PAS cell was used for the three samples. These gases are well-known trace gases in the environment and in exhaled breath. The gases cause environmental degradation through their effects on soil acidification, eutrophication, and stratospheric ozone depletion. The presence of ammonia in the environment are mainly due to the degradation of animal waste, industrial processes and diesel exhaust. NO$_2$ is a very toxic gas and a regulated air pollutant that possess a serious risk for human health. Atmospheric NO$_2$ is primarily a manmade problem and is related to combustion processes, such as the emissions from cars, power plants and factories. The average mixing ratio of NO$_2$ in the atmosphere is typically around 5 parts per billion by volume (ppbV), however it can be orders of magnitude higher close to the pollution source.  Regulations with respect to hazardous gases are becoming stricter and therefore real-time novel sensors with ppbV sensitivity are needed. Methane and ammonia are also biomarkers for breath gas analysis e.g. in identification of inflammation, which may be related to different cancerous diseases or liver failures. We believe that the PA sensor will have the potential to provide a non-invasive method with fast screening capability, thereby significantly improving, early diseases diagnosis for the benefit of patients.

\section{The PA technique}

The PA technique is based on the detection of sound waves that are generated due to the periodic absorption of modulated optical radiation. In the case of pulsed light sources there are two typical modes of operation \cite{Bartlome2009}: The pulse repetition rate is much lower than the resonance frequency of the PA cell and the entire decay curve of the acoustic resonance can be detected and analyzed in a single shot. This is typically achieved using a gated integrator. In this case the generated PA signal is described by a series expansion of decaying sine functions which in the case of high Q cavities has the spectrum with a Lorentzian profiles. On the other hand if the repetition rate can be matched to the resonance frequency of the PA cell, an operation similar to the case of modulated continuous-wave light sources can be achieved. Using the latter approach a PA cell with a relatively high resonance frequency is designed to match the repetion rate of the OPO. A microphone is used to monitor the acoustic waves that appear after the laser radiation is absorbed and converted to local heating via collisions and de-excitation in the PA cell. The voltage magnitude of the measured PA signal is given by \cite{Rosencwaig1980Book}:
\begin{equation}
S_{PA} = S_m P F \alpha,
\label{eq.PAsignal}
\end{equation}
where $P$ is the power of the incident radiation, $\alpha$ is the absorption coefficient, which depends on the total number of molecules per cm$^3$ and the absorption cross section, $S_m$ is the sensitivity of the microphone and $F$ is the cell-specific constant, which depends on the geometry of acoustic cell and the quality factor $Q$ of the acoustic resonance. In the following we present results for spectroscopic measurements of CH$_4$, NH$_3$ and NO$_2$ with PA spectroscopy in the mid-infrared (MIR) region (3.0 $\mu$m to 3.7 $\mu$m) using a pulsed MIR OPO as light source . 

\section{Experimental setup}

\begin{figure}[htbp]
\centering
\fbox{\includegraphics[width=\linewidth]{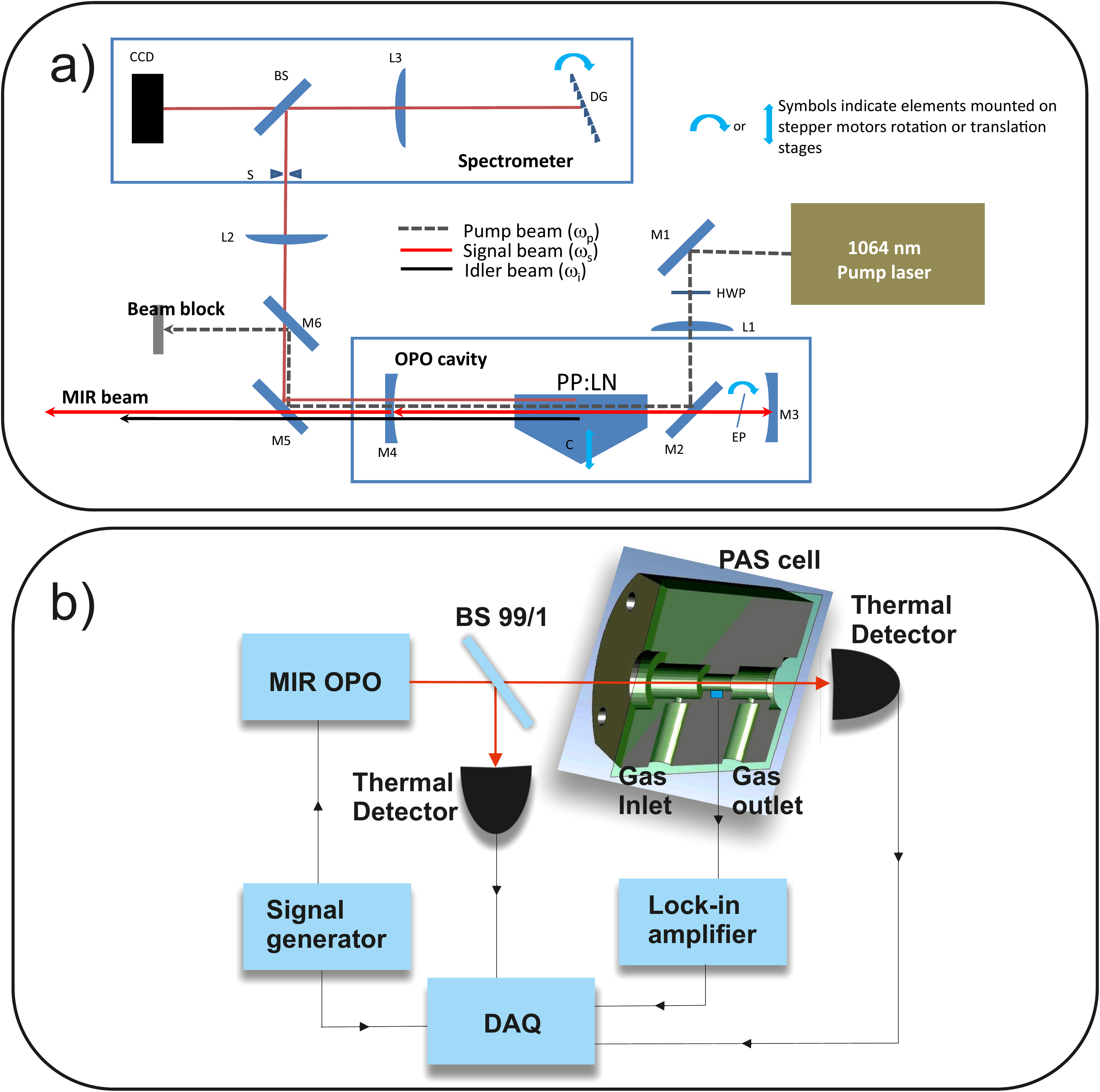}}
\caption{Schematics of the single resonant (NIR signal beam) MIR OPO and illustration of the PAS sensor. a) The MIR OPO is optimized for operation in the spectral region between 3.1 $\mu$m and 3.7 $\mu$m and with an average output power of up to 500 mW and pulse durations of $\sim$ 13 ns at a repetition rate of 13.7 kHz. A computer is controlling the stepping motors where the OPO is attached and provide the data acquisition of the PAS signals. M3 and M4 are high reflectors (HR) with curvature of 100 mm for the signal beam and. M2 is a HR for the 1064 nm pump beam and has high transmittance for the signal and idler beams. b) The PAS cell has a cylinder shaped geometry with a length of 9.5 mm and a diameter of 3 mm with and a fundamental acoustic frequency at approximately 13.7 kHz. }
\label{fig1}
\end{figure}

Figure \ref{fig1} depicts the main components of the PAS sensor with the integrated OPO: Figure \ref{fig1}a) shows the pulsed MIR OPO, and Figure \ref{fig1} b) the complete PAS system. 

\subsection{The MIR OPO}

\begin{figure}[htbp]
\centering
\fbox{\includegraphics[width=\linewidth]{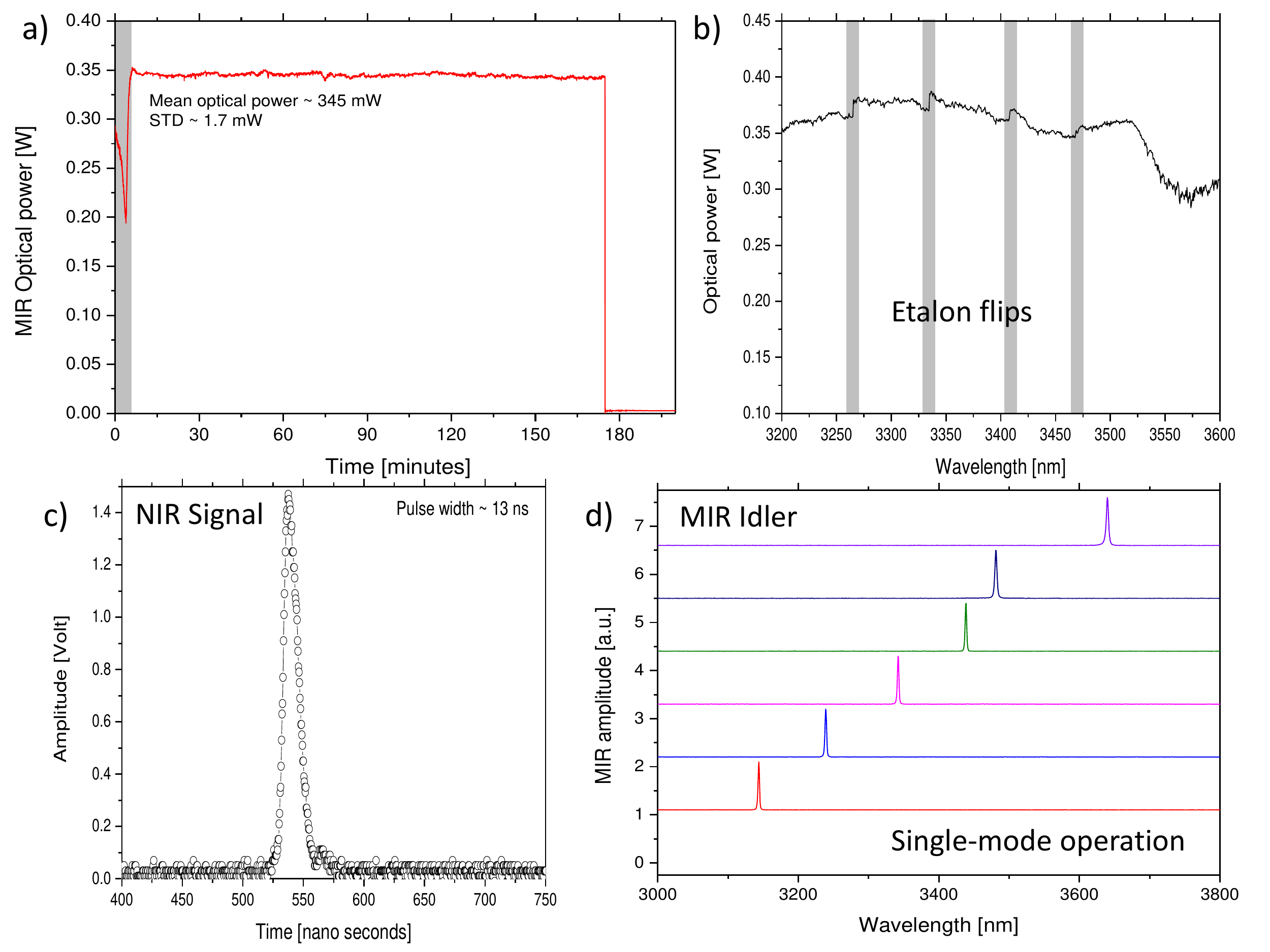}}
\caption{a) The mean optical power as a function of time. During the almost 3h measurement the corresponding standard deviation (STD) was less than 0.5\% RMS. The grey shaded area shows the warm up time for the OPO. b) Continuous spectral tuning of the MIR OPO over 400 nm. The Etalon flips are observed for every 70 nm. c) Pulse characterization of the generated NIR signal beam. Pulse length approximately 13 ns. d) Continuous spectral MIR tuning of the OPO. }
\label{fig1opodata}
\end{figure}

The MIR OPO used is based on a 50 mm long periodically poled lithium niobate (PPLN) nonlinear crystal with a fanned-out structure. The PPLN is placed inside a single-resonant cavity, resonant for the near infrared (NIR) signal beam. The OPO is side pumped at 1064 nm with diode-pumped nanosecond laser having an average output power reaching up to 15W. The Q-switch repetition rate can be changed continuously from 10 kHz to 80 kHz and the pulse duration from 7 to 50 ns. The continuously tuning of the repetition rate makes it possible to synchronize the repetition rate of the OPO with the acoustic resonance of the PAS cell, thus being equivalent to a modulated continuous-wave light PAS setup and the PAS signal is then simply processed with a lock-in amplifier. The maximum wavelength ranges for the NIR signal and MIR idler are 1.4 $\mu$m to 1.7 $\mu$m and 2.8 $\mu$m to 4.6 $\mu$m, respectively, and with a MIR output power up to 500 mW. The pulse peak power in this case is approximately 2.8 kW. The tunability of the OPO is achieved by the vertical translation of the fanned-out PPLN crystal, while the temperature of the crystal is kept at 30$^\circ$C. The system is kept single-mode using a 50$\mu$m thick YAG etalon plate. This provides a bandwidth of around 1-1.5 cm$^{-1}$. In order to ensure single longitudinal mode (SLM) operation, the sum frequency generated beam of the pump and signal is monitored with a spectrometer that is integrated in the system. A computer controls the grating, etalon plate and the stepping motors for nonlinear crystal position. Figure \ref{fig1opodata} shows the performance data of the MIR OPO. Figure \ref{fig1opodata}a shows the power stability of the OPO when operating at 3.4 $\mu$m performs almost 3h of measurement. The mean optical power was 345 mW and corresponding standard deviation (STD) was less than 0.5\% RMS. Figure \ref{fig1opodata}b shows the mean MIR optical power as function of wavelength. We observe that the OPO can be tuned approximately 70 nm before the etalon is flipped, which gives rise to a small power fluctuation as seen in figure \ref{fig1opodata}b. The optical power is varying slightly over the tuning range and this needs to be taken into account when performing spectroscopic PA measurements by normalizing the PA signal with the optical power. Figure \ref{fig1opodata}c shows the pulse characterization of the generated NIR signal beam with a pulse length of $\approx$ 13 ns (FWHM). In figure \ref{fig1opodata}d the continuous spectral MIR tuning of the OPO is shown. The spectra show that the OPO operates in single-mode operation in the whole lasing area of interest. The spectra were measured using an optical spectrum analyser with a resolution of 0.3 nm@3.5$\mu$m. 

\subsection{The PA cell}

\begin{figure}[htbp]
\centering
\fbox{\includegraphics[width=\linewidth]{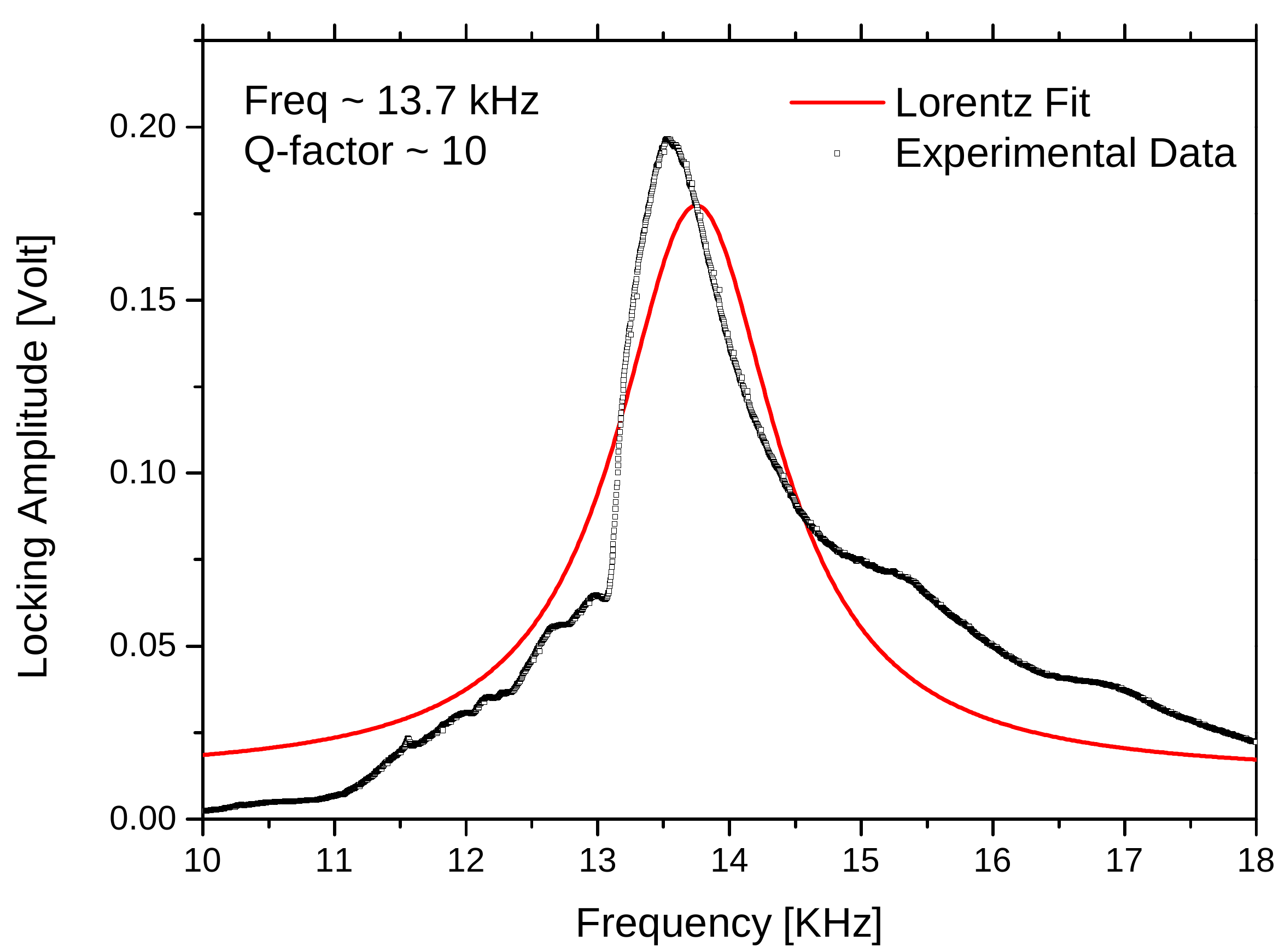}}
\caption{Frequency response of the PAS cell. The black curve is the experimental data and the red curve is a Lorentz fit. The resonance frequency was found to be at 13.7 kHz with a Q-factor of approximately 10. }
\label{Fig_resonance}
\end{figure}

A typical setup for PAS includes an amplitude or frequency modulated light source and an absorption cell with microphones (or any pressure sensitive element), most PAS systems are operated resonantly, thus the PA signal is enhanced with a factor that is proportional to the Q-factor of the acoustic resonances \cite{Rosencwaig1980Book}. In the present work the OPO is optimized for operation in the spectral region between 2.8 $\mu$m and 3.8$\mu$m, with an average output power of 345 mW and pulse durations of $\sim$ 13 ns at a repetition rate of 13.7 kHz. The laser beam enters and exits the PA cell via uncoated 3 mm thick calciumfluorid windows. The optical transmission is approximately 90$\%$ and the optical power is monitored by a thermal detector before and after the PAS cell. The PAS cell is made out of Polyoxymethylene (POM). POM is an engineering thermoplastic used in precision parts requiring high stiffness, low friction, and excellent dimensional stability. The PAS cell has cylinder shaped geometry with a length of 9.5 mm and a diameter of 3 mm. The flow buffer zones have length of 18 mm and a diameter of 8 mm. The gas inlet and outlet are placed in the buffer zones in order to reduce the coupling of external acoustic noise to the acoustic resonator. The $f$ frequency of the first longitudinal resonance mode for an open cylindrical resonator can be found from $f=c/(2l+1.6d)$, where $c$ is the speed of sound of the gas inside the resonator, $l$ is the length of resonator and $d$ is diameter of the cell. The speed of sound in dry air at 20$^o$C is 343 m/s, thus the first longitudinal mode is expected to have its frequency at $\sim$ 14.4 KHz. By designing the PA cell with this relatively high resonance frequency the PA cell can be excited resonantly with the MIR OPO and the operation of the PA sensor is equivalent to the case of using modulated continuous-wave light sources for PAS \cite{Bartlome2009}. Figure \ref{Fig_resonance} shows the measured acoustic resonance of the PAS cell. Experimentally the resonance frequency was found to be at 13.7 kHz and with a Q-factor of approximately 10. The volume of the absorption cell is 0.067 ml, thus using a constant gas flow rate of up to 300 ml/min the gas is exchanged more than 4000 times pr. minute ensuring fast gas sampling. The microphone is placed in the middle of the absorption tube, where the acoustic sound pressure is expected to have the highest amplitude for the fundamental acoustic mode. The voltage signal from the microphone is amplified by a pre-amplifier with a 10-20 kHz bandpass filter before being processed with a lock-in amplifier and finally being digitized with a 12-bit data acquisition (DAQ) card.
\section{PAS results}

\begin{figure}[htbp]
\centering
\fbox{\includegraphics[width=\linewidth]{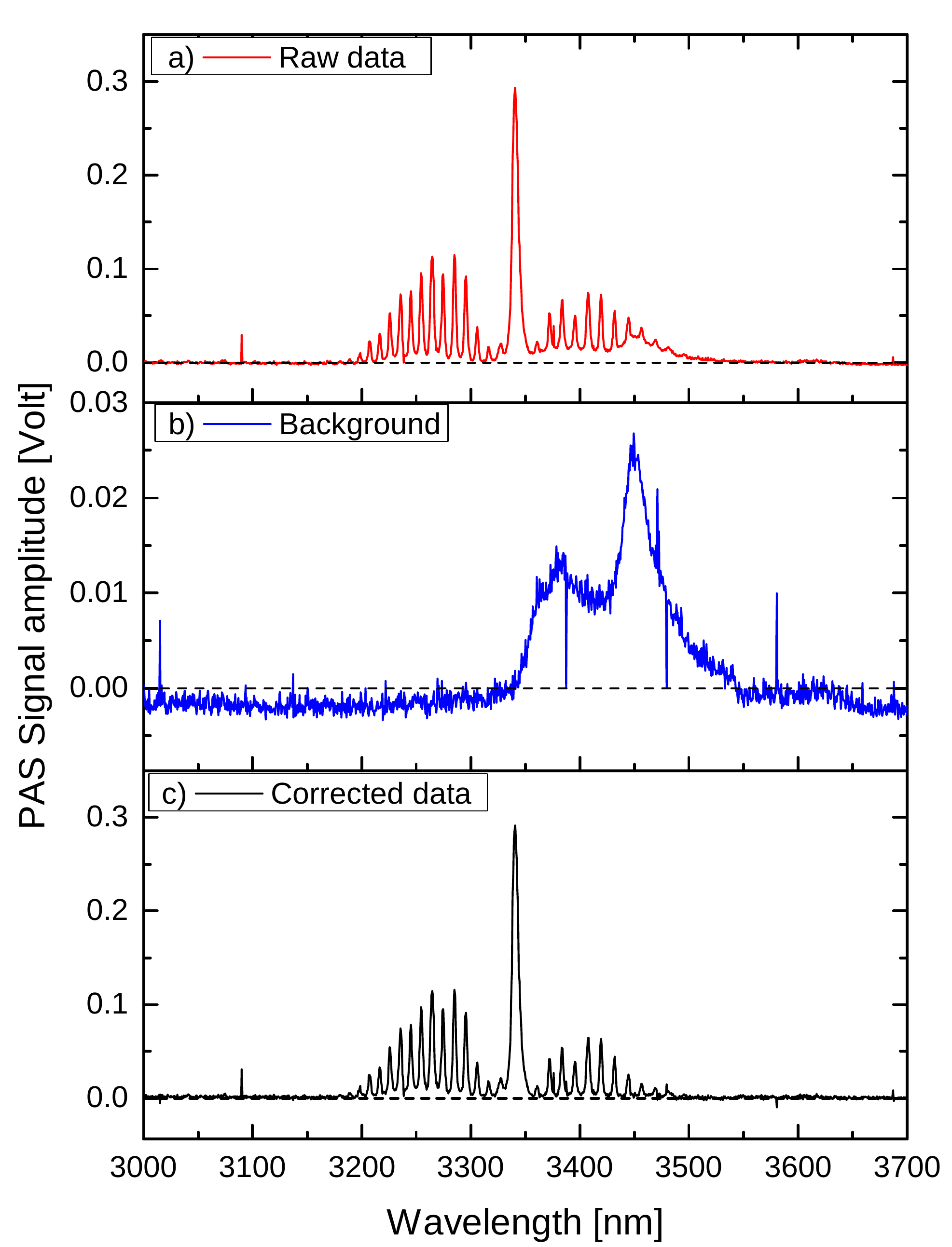}}
\caption{Spectral PAS measurements. a) raw data of 100 ppmV CH$_4$ in N$2$ without any corrections, b) background signal when flowed with nitrogen and c) 100 ppmV CH$_4$ in N$2$ corrected for the background signal seen in b). The correction is simply archieved by subtracting the background from the spectra.}
\label{fig5}
\end{figure}

The PAS experiments were performed by excitation of molecular ro-vibrational transitions of 100 ppmV of methane in nitrogen in the 3.0-3.6 $\mu$m wavelength region. Figure \ref{fig5} shows the measured spectrum of methane (CH$_4$) with a 100 ppm concentration in nitrogen. We clearly see the R-, Q- and P-branch of methane. The data were processed with a lock-in amplifier with a time constant of 100 ms. The wavelength of the OPO was changed in steps of 0.5 nm. All measurements were made with constant gas flow of 300 ml/min.

\begin{figure}[htbp]
\centering
\fbox{\includegraphics[width=\linewidth]{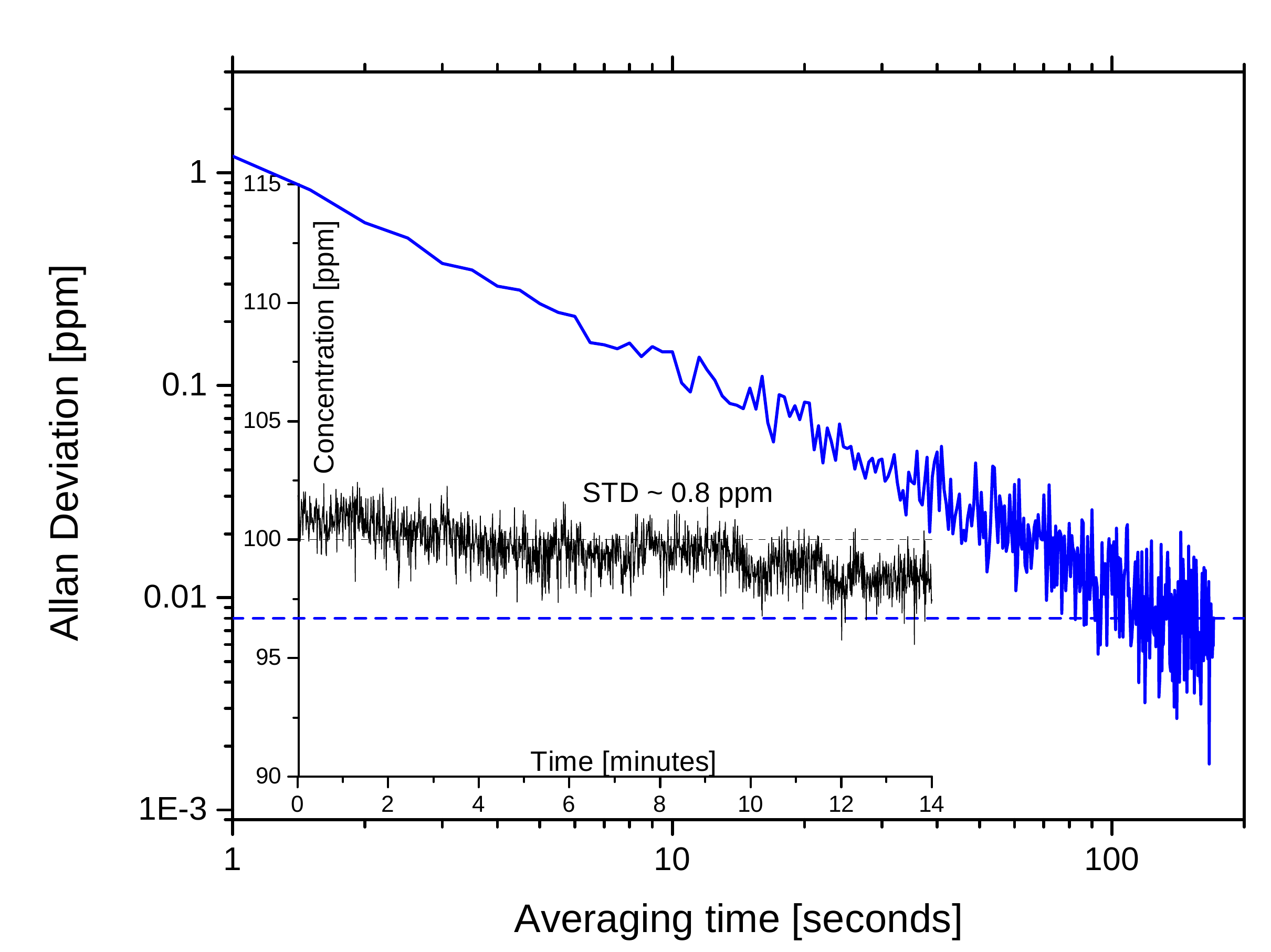}}
\caption{Allan deviation analysis of the lock-in signal with 1 second integration time and measured at 3.32 $\mu$m. The inset figure is the measured time trace, while the cell is flowed with 100 ppm of CH$_4$. Detection sensitivity at optimum integration time is 8 ppbv@105.}
\label{fig6}
\end{figure}

\begin{figure}[htbp]
\centering
\fbox{\includegraphics[width=\linewidth]{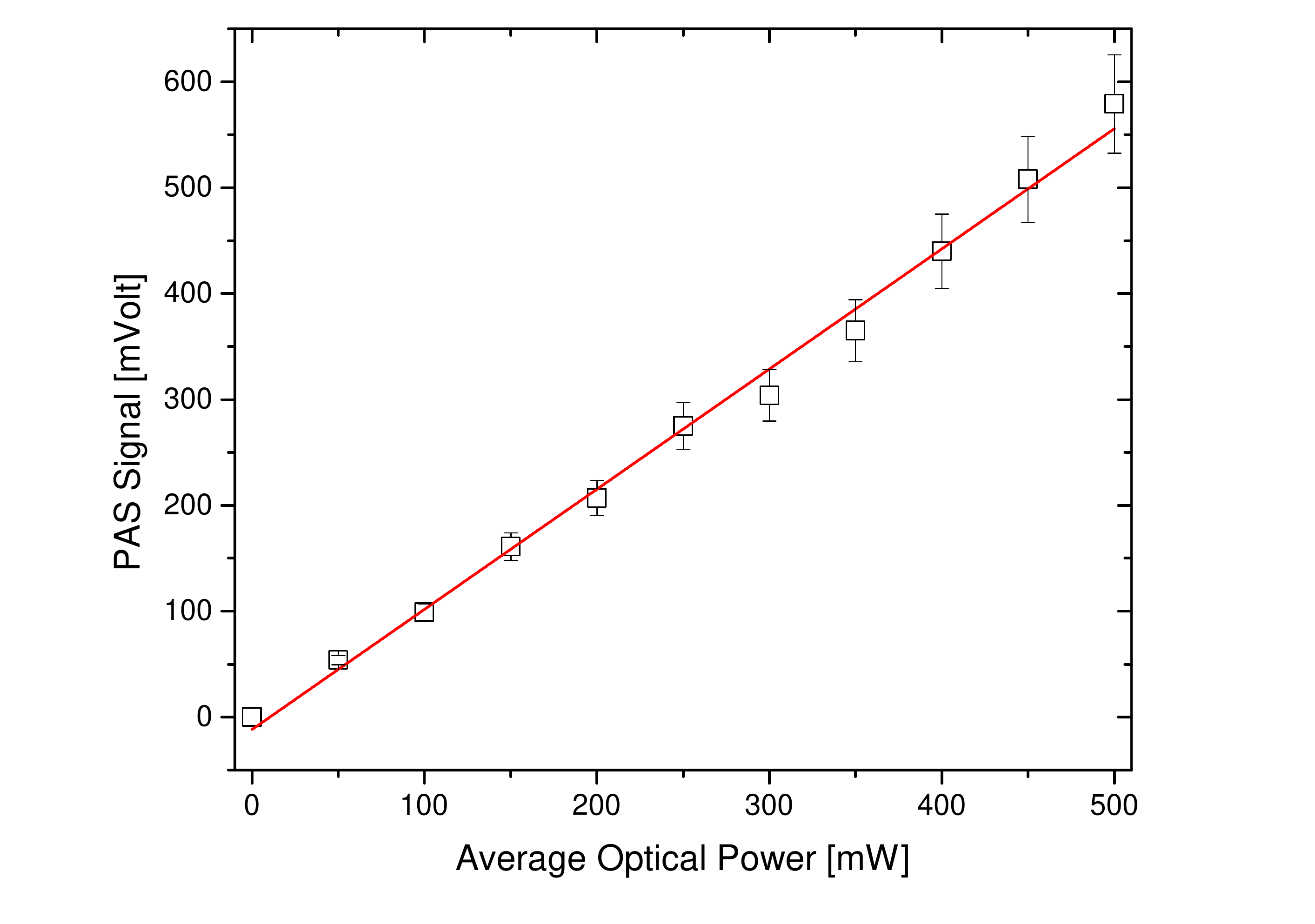}}
\caption{PA signal as function of pump power. The error bars are 8$\%$. The linear regression is 0.99. }
\label{fig7}
\end{figure}

In order for the trace gas analyser platform to be practical, having a small resolution bandwidth and high tunability is not sufficient, the sensor also needs high sensitivity in the range ppbV \cite{Yin2017,lassen2014,Peltola2015,Szabo2013}. We find that the single shot noise equivalent detection sensitivity (NEDS) is 0.8 $\pm 0.1$ ppm. The NEDS describes the PAS sensor performance on a short time scale, however, characterization of long-term drifts and signal averaging limits is very important for sensors. There exists an optimum integration time at which the detection limit reaches a minimum value. At longer averaging time, drift effects emerge and the sensor performance deteriorates. The optimum integration time and detection sensitivity are therefore determined using an Allan deviation analysis. The OPO was locked to the absorption peak at 3.42 $\mu$m while the cell is flowed with 100 ppmV of CH$_4$. The inset in Figure \ref{fig6} shows the recorded time trace during 14 minutes and thus constitute the time trace of the 100 ppmV CH$_4$ measurement. Figure \ref{fig6} shows the Allan deviation analysis for the PAS data processed with a lock-in amplifier with an integration time of 1 second. The analysis shows that the detection sensitivity at optimum integration time is 8 ppbv@105s for CH$_4$ measured at the Q-peak of CH$_4$ at 3.32 $\mu$m, corresponding to a normalized noise equivalent absorption (NNEA) coefficient of 2.9$\times 10^{-7}$ W cm$^{-1}$ Hz$^{1/2}$. In figure \ref{fig7} we show measurements of the PA signal as a function of the average optical pump power. Again the OPO was locked to the Q-branch peak of methane at 3.42 $\mu$m while the cell is flowed with 100 ppmV of CH$_4$. In the optimized working range 100-500 mW the PA sensor has a linear dependence to the optical power as also expected from equation \ref{eq.PAsignal}. However, saturation of optical absorption might occur above a certain laser pump power level and therefore the PA signal might not increase linearly with increasing laser power.

\begin{figure}[htbp]
\centering
\fbox{\includegraphics[width=\linewidth]{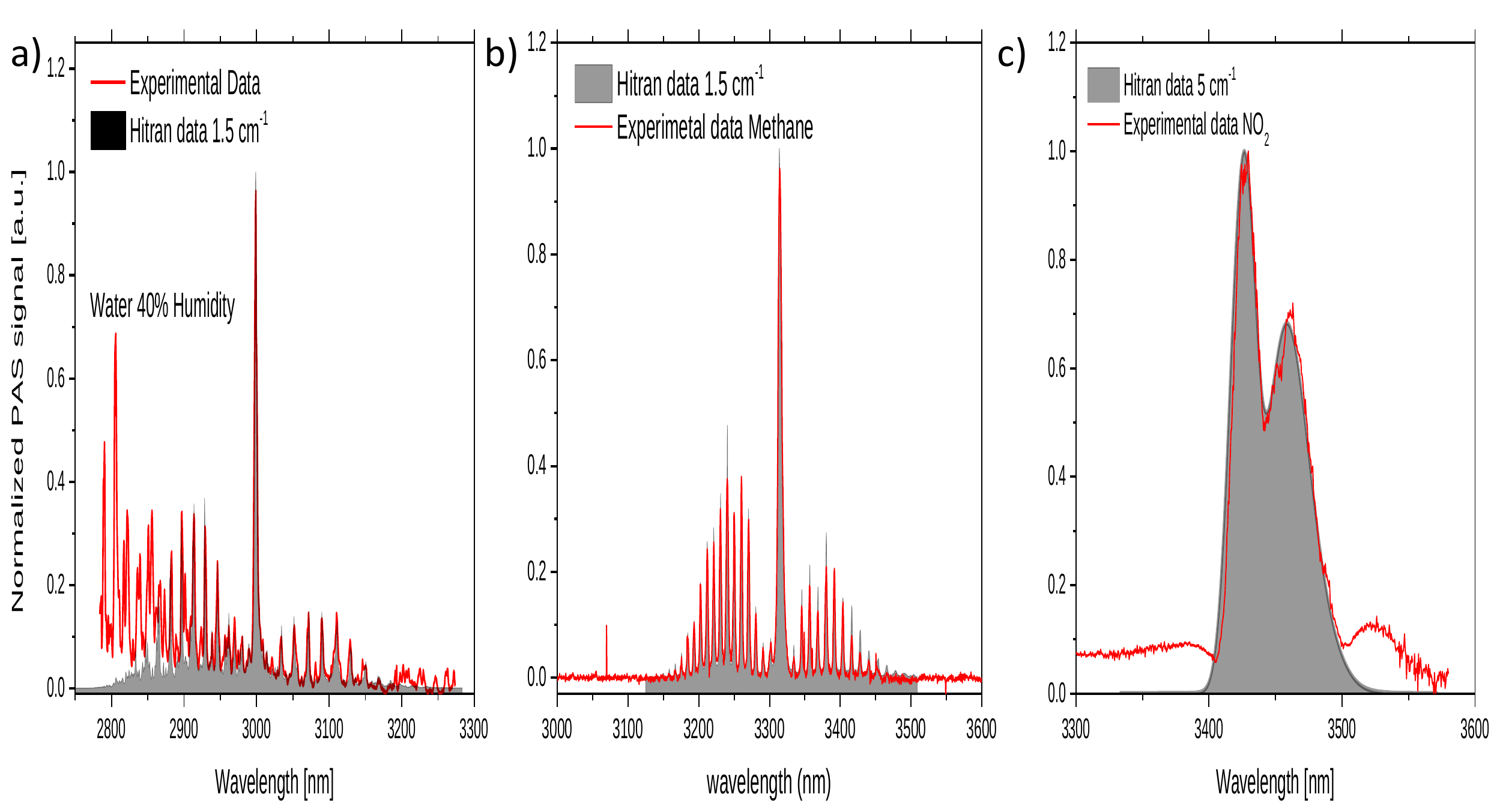}}
\caption{Correct spectral PAS measurements. a) The red shows the measured ammonia spectrum with an estimated 1000 ppmV $\pm$ 50 ppmV concentration and at 1 atm pressure. The concentration is estimate from the methane PAS signal seen in Figure \ref{fig5}. 40$\%$ humidity (water) can also been seen and differentiated from the ammonia spectrum. b) 100 ppmV CH$_4$ in N$2$ and c) 100 ppmV NO$_2$ in N$2$ }
\label{fig8}
\end{figure}

Figure \ref{fig8} shows the background corrected spectral PAS measurements of a) ammonia (NH$_3$),b) methane (CH$_4$) and c) nitrogen dioxide (NO$_2$). The correction is simply achieved by subtracting the background from the spectra. For the fitting of the ammonia and methane data with the Hitran data we have used a convolution of 1.5 cm$^{-1}$ as shown in figure \ref{fig8}a) and b), thus we estimate the spectral resolution of our PAS sensor to be 1.5 cm$^{-1}$. Note that a nanosecond pulsed tuneable MIR OPO has previously been used for PAS measurements of methane, however, only in a narrow wavelength range \cite{Hellstrom2013}. Our system is caple of scanning much broader wavelength ranges and therefore more versatile for many different trace gases with the same setup. This is shown in the following where measurements of water and ammonia is demonstrated in the 2.8-3.2 $\mu$m range. The PAS technique is not an absolute technique calibration is required against a known gas sample concentration. However, when the PAS sensor is calibrated we can use the measured methane signal for the calibration of the PAS sensor. In this case we use the obtained data for methane to estimate the concentration of an unknown amount of ammonia. Figure \ref{fig8}a) shows the measured spectrum of ammonia (NH$_3$). The grey curve shows the measured ammonia spectrum at 1 atm pressure. The red curve shows data from the Hitran database. The R-, Q- and P-branch of ammonia is clearly observed. We estimate the ammonia concentration to be approximately 1000 ppmV $\pm$ 50 ppmV. In the obtained spectrum a humidity (water) of 40$\%$ can also been seen and clearly differentiated from the ammonia spectrum. Note that for breath gas analysis ammonia is an interesting biomarker, since it has been indicated that ammonia is a biomarker for different liver metabolism diseases and liver malfunctions . Finally we have conducted spectroscopic measurements of NO$_2$ in the 3250 nm to 3550 nm wavelength region with a resolution bandwidth of 5 cm$^{-1}$ and a single-shot detection limit of 1.6 ppmV is demonstrated. The acquired spectrum is compared with data from the Hitran database and a very good agreement is found. The reason for the slightly worse spectral resolution is due to the fact that the integration time of the lock-in amplifier was increased to 1 second.

\section{Conclusion}
We have demonstrated a miniaturized PAS configuration  system pumped resonantly by a nanosecond pulsed single-mode MIR OPO. Different spectral branches of methane (CH$_4$) and ammonia (NH$_3$) are resolved and clearly identified and compared with the HITRAN database. From the comparison of the spectra with the Hitran spectra we conclude that the presented PAS sensor resolution bandwidth is approximately 1.5 cm$^{-1}$ and is limited by the thickness of the etalon plate and the integration time used. The spectroscopic methane measurements have a detection sensitivity of approximately 0.8 ppmV. However by applying optimum integration time the sensitivity can be improved to 8 ppbV ($\mu$mol/mol) at 105 seconds of averaging, corresponding to a normalized noise equivalent absorption (NNEA) coefficient of 2.9$\times 10^{-7}$ W cm$^{-1}$ Hz$^{1/2}$. We believe that the tunability and sensitivity demonstrated here will make the PAS sensor very useful for future environmental measurements and breath gas screenings, due to its ease of use, compactness , fast response time and its capability of allowing trace gas measurements at the sub-parts per billion (ppb) level. The sensor will be capable of detecting early stages of diseases almost instantaneously providing data for potential diagnostics.

\section{Funding Information}
We acknowledge the financial support from EUREKA (Eurostars program: E10589 - PIRMAH) and the Danish Agency for Higher Education.

\end{document}